\documentclass[reprint,twocolumn,amsmath,amssymb,aps,prl,superscriptaddress]{revtex4-1}
\usepackage{graphicx}
\usepackage{dcolumn}
\usepackage{bm}
\usepackage{float}
\usepackage{amssymb}
\usepackage{color}
\usepackage{multirow}
\usepackage{stmaryrd}
\usepackage{sidecap}

\begin{document}

\title{Quantum oscillations in the non-centrosymmetric superconductor and topological nodal-line semimetal PbTaSe$_2$}

\author{Xitong Xu}
\affiliation{International Center for Quantum Materials, School of Physics, Peking University, China}
\author{Zhibo Kang}
\affiliation{International Center for Quantum Materials, School of Physics, Peking University, China}
\author{Tay-Rong Chang}
\affiliation{Department of Physics National Cheng Kung University Tainan 701, Taiwan}
\affiliation{Center for quantum frontiers of research \& technology (QFort)}
\author{Hsin Lin}
\affiliation{Institute of Physics, Academia Sinica, Taipei 11529, Taiwan}
\author{Guang Bian}
\affiliation{Department of Physics and Astronomy, University of Missouri, Columbia, Missouri 65211, USA}
\author{Zhujun Yuan}
\affiliation{International Center for Quantum Materials, School of Physics, Peking University, China}
\author{Zhe Qu}
\affiliation{Anhui Province Key Laboratory of Condensed Matter Physics at Extreme Conditions, High Magnetic Field Laboratory of the Chinese Academy of Sciences, Hefei 230031, Anhui, People's Republic of China}
\author{Jinglei Zhang}
\email{zhangjinglei@hmfl.ac.cn}
\affiliation{Anhui Province Key Laboratory of Condensed Matter Physics at Extreme Conditions, High Magnetic Field Laboratory of the Chinese Academy of Sciences, Hefei 230031, Anhui, People's Republic of China}
\author{Shuang Jia}
\email{gwljiashuang@pku.edu.cn}
\affiliation{International Center for Quantum Materials, School of Physics, Peking University, China}
\affiliation{Collaborative Innovation Center of Quantum Matter, Beijing 100871, China}
\affiliation{CAS Center for Excellence in Topological Quantum Computation, University of Chinese Academy of Sciences, Beijing 100190, China}
\affiliation{Beijing Academy of Quantum Information Sciences, West Bld.\#3, No.10 Xibeiwang East Rd., Haidian District, Beijing 100193,China}

\begin{abstract}
We observed quantum oscillations in thermoelectric and magnetic torque signals in non-centrosymmetric superconductor PbTaSe$_2$.
One oscillatory frequency stems from the orbits formed by magnetic breakdown, while others are from two-dimensional-like Fermi surfaces near the topological nodal rings.
Our comprehensive understanding of the Fermi surface topology of PbTaSe$_2$, including nailing down the Fermi level and detecting the Berry phases near the nodal rings, is crucial for searching plausible topological superconductivity in its bulk and surface states.

\end{abstract}

\date{\today}
\maketitle

\section{Introduction}

Searching topological superconductors (TSCs) in real materials has been an exciting endeavor in condensed matter physics, as they are closely related to Majorana fermion which can be used to realize topological protected quantum computation\cite{annu-fuliang-TSC,Rep-Sato-TSC}.
One kind of TSC candidates are non-centrosymmetric superconductors (NCSCs) with strong spin-orbit coupling (SOC). Because of the breaking of spin degeneracy by asymmetric SOC, these NCSCs can manifest a parity-mixed
superconducting state\cite{smidman2017superconductivity} whose edge state may host Majorana fermions if the $p$-wave gap is larger than $s$-wave gap\cite{PhysRevB.79.094504, PhysRevB.79.060505}.
The research on the NCSCs with strong SOC\cite{yuan2006s} has presented a significant direction since the discovery of the unconventional heavy-fermion NCSC\cite{PhysRevLett.92.027003,bauer2004novel,bauer2012non}.

Recently, a layered compound PbTaSe$_2$ was found to be superconducting with T$_c$ around 3.8 K\cite{ali2014noncentrosymmetric,wang2016nodeless,zhang2016superconducting}. It displays a highly non-centrosymmetric structure where triangle lattices of Pb atoms are sandwiched between hexagonal TaSe$_2$ layers. The heavy elements in PbTaSe$_2$ induce strong SOC, and thereby a large Rashba splitting\cite{ali2014noncentrosymmetric}. Specific heat measurements\cite{zhang2016superconducting} reveal a full superconducting gap with no gapless nodes, indicating the triplet state is not dominant.
Other experimental evidences, including the field dependence of residual thermal conductivity\cite{wang2016nodeless}, the upward curvature of the upper critical field\cite{wang2015upward}, together with $^{207}$Pb nuclear magnetic resonance measurements\cite{wilson2017mu} suggest a scenario of multiband superconductivity in PbTaSe$_2$.

Interestingly, angle-resolved photoemission spectroscopy (ARPES) measurements and quasi-particle scattering interference imaging, together with first principle calculations, reveal the existence of bulk nodal-line band structure and fully spin-polarized topological surface states (TSSs) in PbTaSe$_2$\cite{bian2016topological,chang2016topological,guan2016superconducting}.
The existence of the TSSs points to another possibility towards a surface TSC which is induced by proximity effect through the bulk $s$-wave superconductor\cite{hosur2011majorana,fu2008superconducting}.

The TSSs in PbTaSe$_2$ emanate from surface projection of the topological nodal rings (TNRs) which are generated by the asymmetric SOC and protected by the reflection symmetry.
As these nodal rings are all located above the Fermi level ($E_F$), they have yet to be directly confirmed by ARPES experiments.
Moreover, due to the limited resolution of ARPES, the fine structures related to the TNRs near the Fermi surface (FS) have not been identified by experiment.
Here we present the first investigation of the complicated, fine structures of the TNRs via analysing quantum oscillations (QOs) in magneto-thermopower and magnetic torque signals.
Combining band structure calculations, we precisely depict the SOC split TNRs and the exact $E_F$ location.
Moreover, we detected a nontrivial Berry phases of electron orbits interlocking with the TNRs.
Our finding confirms the existence of the TNRs at the exact $E_F$ in this NCSC, allowing better understanding of possible topological superconductivity in its bulk and surface states.

\section{Method}

High-quality single crystals of PbTaSe$_2$ in our studies were synthesized by standard chemical vapor transport method\cite{zhang2016superconducting, 2015arXiv151105295S}.
Thermopower measurement was carried out in a 14 T Oxford Teslatron PT system, using a one-heater-three-thermometer setup in which the temperature gradient is applied in one sample (labeled as P1) along the crystallographic $a$ direction and magnetic field $\bm{H}$ along the $c$ axis.
The voltage signals were amplified using EM DC Amplifier A10 and subsequently collected in a Keithley 2182A nano-voltmeter.
With careful setting-up, the peak-to-peak noise level in our system is less than 2 $\mathrm{nV}$.
N. B. a small voltage offset ($\sim -0.2~\mathrm{\mu V/K}$ at 2 K) contributed from reverse temperature gradient in the manganin leads\cite{rathnayaka1985thermoelectric} has been subtracted.
Magnetic torque measurement was performed in Chinese High Magnetic Field Laboratory (CHMFL) in Hefei using a resistive water-cooled magnet in fields up to 33 T on the other sample (labeled as P2).
The torque signal was detected via conventional CuBe capacitance cantilever\cite{zhang2018non}.
The device is fixed on a platform that could be rotated \it in situ \rm around one axis.
Band structure calculations were performed under the framework of the generalized gradient approximation of density functional theory (DFT)\cite{PhysRevLett.77.3865} as implemented in the VASP package\cite{kresse1996efficiency}. The Fermi surface in Fig.~\ref{f2}(c) and (d) is generated by XCrySDen\cite{kokalj2003computer}.

\section{Experimental Results and Discussion}

\begin{figure}[htbp]
\begin{center}
\includegraphics[clip, width=0.49\textwidth]{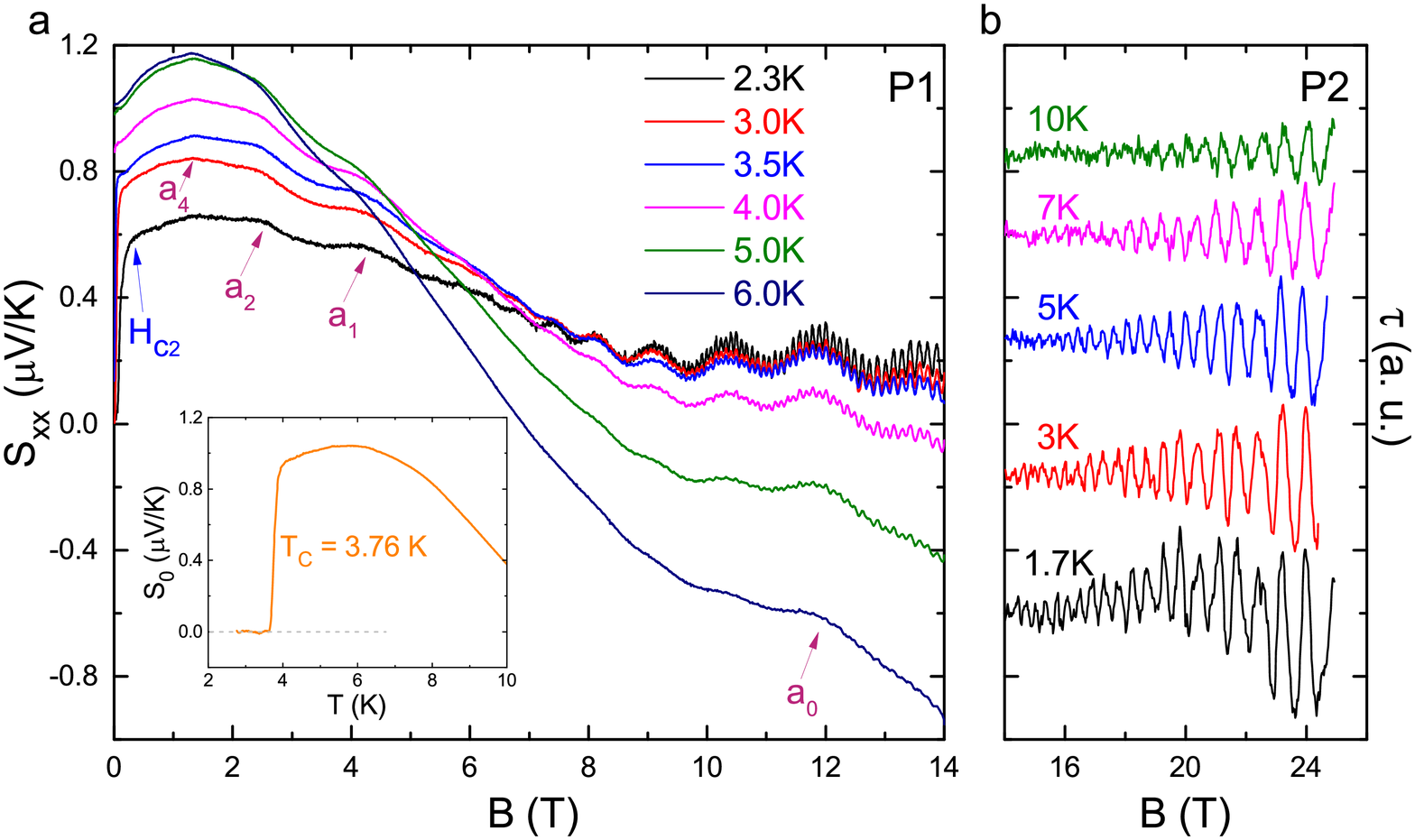}  
\caption{Color online. (a) Magneto-thermopower of PbTaSe$_2$ (sample P1) at selected temperatures. Purple arrows indicate the positions of the oscillatory peaks corresponding to the lowest oscillating frequency ($\sim$ 6 T) and  $a_0$ to $a_4$ denote the corresponding Landau indexes. The blue arrow points to the upper critical field of superconducting transition. Inset: zero-field Seebeck coefficient of PbTaSe$_2$ versus temperature.   (b) Magnetic torque signals for sample P2 at selected temperatures in higher magnetic fields. Curves are offset for clarity. }
\label{f1}
\end{center}
\end{figure}

The inset in Fig.~\ref{f1}(a) shows zero-field Seebeck coefficient ($S_0$) of PbTaSe$_2$ at low temperature.
A sharp superconducting transition is apparent in $S_0$ at T$_c$ = 3.76 K, which is consistent with previous results in resistivity and magnetic susceptibility\cite{ali2014noncentrosymmetric,wang2016nodeless,zhang2016superconducting}.
When temperature is higher than T$_c$, $S_0$ restores to a finite positive value of around $1~ \mathrm{\mu V/K}$. At about 6 K, $S_0$ begins to decrease and changes its sign at higher temperature.
This feature agrees with the existence of multiple types of carriers in PbTaSe$_2$.
Magneto-Seebeck signals ($S_{xx}$) at different temperatures are shown in Fig.~\ref{f1}(a). Above the upper critical field H$_{c2}$, $S_{xx}$ firstly increases and then bends down above 1.4 T.
Because the carrier density in PbTaSe$_2$ is of the order of $5\times 10^{21}~\mathrm{cm}^{-3}$ (estimated from Hall resistivity), the response of $S_{xx}$ to field is small.
Yet strong QOs with multiple frequencies are apparent.
A series of oscillatory peaks with a very low frequency (6 T) are indicated by arrows in Fig.~\ref{f1}(a).
Above 6 T, QOs with a much higher frequency is modulated on another low frequency, and both sets of QOs damp rapidly with increasing temperature.
Significant QOs were also observed in torque signal $\bm{\tau}=\bm{M}\times\bm{B}$ arising from the magnetic susceptibility anisotropy.
As shown in Fig.~\ref{f1}(b), QOs in torque signals also become apparent after subtracting a proper background when the field exceeds 14 T.

\begin{SCfigure*}
\centering
\includegraphics[clip, width=0.72\textwidth]{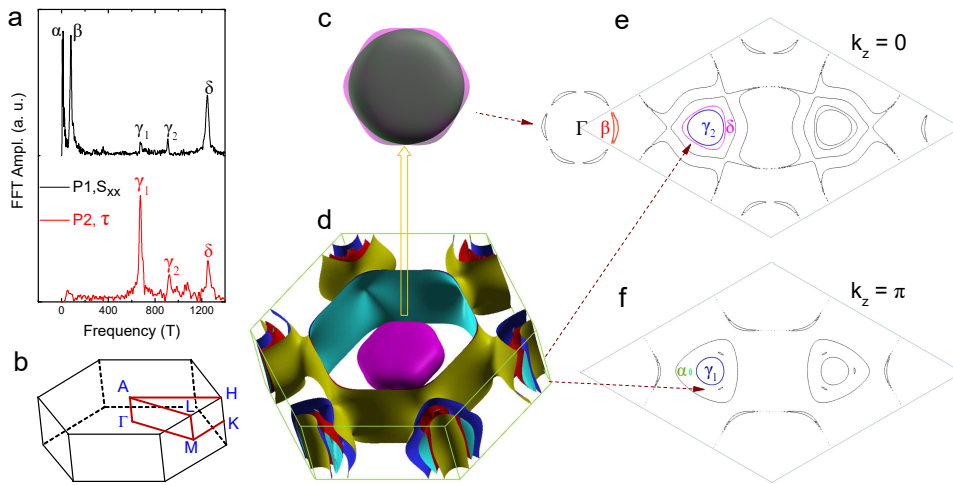}  
\hspace{.02\textwidth}
\caption{Color online. (a) FFT spectrum of the observed QOs in $S_{xx}$ and $\tau $.   (b) A sketch of the first BZ for PbTaSe$_2$.  (c) FS of PbTaSe$_2$ around $\Gamma$ point. The outer FS is set to be translucent to reveal the inner nesting discus-like one.   (d) Overall FS of PbTaSe$_2$ in the first BZ.   \ (e) and (f) Cuts of FS at $k_z = 0$ and $\pi$, respectively. N. B. these two cuts are plotted in the primitive cell of the 2D reciprocal space in order to show closed orbits. Colored orbits correspond to the frequencies in the FFT spectrums.}
\label{f2}
\end{SCfigure*}

In order to compare the results of thermoelectric and magnetic torque measurements, we performed fast Fourier transformation (FFT) on the oscillatory parts of $S_{xx}$ and $\tau$.
As shown in Fig.~\ref{f2}(a), five distinct frequencies, which range from 6 T to 1250 T, can be seen on the FFT spectrum of $\Delta S_{xx}$.
We labeled them as $\alpha$, $\beta$, $\gamma_1$, $\gamma_2$ and $\delta$, respectively.
On the FFT spectrum of $\tau$,  only the latter three can be clearly distinguished.
Lower oscillation frequencies in $\tau$ are sensitive to the field window and subtracted background, and therefore we are not able to detect them in high field.
Regardless of the different characterization on different samples, the frequencies of $\gamma_1$, $\gamma_2$ and $\delta$ match well in $S_{xx}$ and $\tau$.

\begin{figure}[htbp]
\begin{center}
\includegraphics[clip, width=0.48\textwidth]{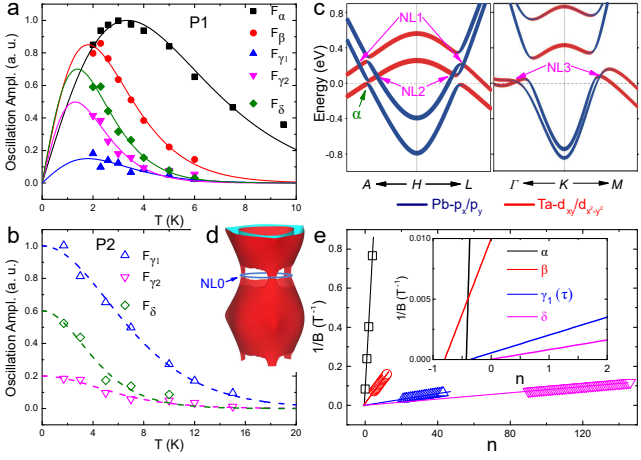}  
\caption{Color online. (a) Fitting of the effective mass ($m^\ast$) using the derivative of LK formula for the five frequencies detected in the QOs of magneto-thermopower of P1. The amplitude of frequency $F_\alpha$ is fitted by using the oscillation at 4.25 T while the others are from the amplitude of corresponding FFT peaks.  (b) Fitting of $m^\ast$ with standard LK formula for torque signals.  (c) Band dispersion around $H$ and $K$ points with SOC. There are two nodal-lines (NL1 and NL2) surrounding $H$ in the $k_z = \pi$ plane and one more (NL3) around $K$ in the $k_z = 0$ plane. The intersections of the nodal rings along the high-symmetry lines are marked with purple arrows. The position of the $\alpha$ pocket is marked green. (d) The torus-like hole pocket plotted in extended BZ. Blue circle represents the position of the spinless nodal ring (NL0) without SOC, where the $\alpha$ orbit also resides.  (e) Landau fan diagrams for $\alpha$, $\beta$, $\delta$ bands subtracted from $S_{xx}$ and $\gamma_1$ from $\tau$.  Inset: zoom-in fitting near $n = 0$ which shows residual phases.}
\label{f4}
\end{center}
\end{figure}

According to the Onsager relation\cite{shoenberg2009magnetic}, every QO frequency $F$ is related to an extremal cross-section area $A_k$ of the electron/hole pockets in momentum space where $A_k=(2\pi e / \hbar) F$.
The observed frequencies correspond to a fraction varying between only 0.014\% to 3.1\% of the basal plane area of the first Brillouin zone (BZ).
The smallest orbit $\alpha$ is comparable in size to that of prototype Weyl semimetal TaAs\cite{PhysRevB.95.085202,zhang2016signature}, while the $\gamma_1$, $\gamma_2$ and $\delta$ orbits are of similar order to the large 2D-like Fermi pockets in nodal-line semimetal ZrSiS\cite{matusiak2017thermoelectric,hu2017nearly,pezzini2018unconventional}.
We then compare our experimental observations with the DFT calculated complicated FS of PbTaSe$_2$ (Fig.~\ref{f2}(c) and (d)).
There are two three-dimensional (3D) hole pockets centered around the BZ center $\Gamma$ in a Russian-doll structure.
The inner one is discus-like, and touches the outer rounded hexagonal-prism-shaped pocket in every $A-L-M-\Gamma$ plane.
Surrounding the two 3D pockets, there are also a pair of quasi-2D hexagon-shaped cylinders around $\Gamma$, which connect to the strongly corrugated sub-branches centered on the zone corner $K-H$ line. The innermost sub-branch $\gamma$ is a pear-shaped cylinder-like electron pocket around the $K-H$ line, while the second-inner one is torus-like hole pocket which branches to tiny separated rods near $k_z=\pi$ plane (also shown in Fig.~\ref{f4}(d)).

To show the extremal closed orbits at $k_z=0$ and $\pi$, we plot the Fermi contours in the primitive cell in the 2D reciprocal space in Fig.~\ref{f2}(e) and (f), respectively.
The two hole pockets around $\Gamma$ adjoin with each other in every $A-L-M-\Gamma$ plane.
This leads to a magnetic breakdown process\cite{shoenberg2009magnetic} when magnetic field is on, forming six crescent-like $\beta$ orbits in the $k_z=0$ plane.
The 2D pear-shaped electron pocket contributes a belly orbit $\gamma_2$ and a neck orbit $\gamma_1$ around $K$ in the $k_z=0$ plane and $H$ in the $k_z=\pi$ plane, respectively.
The torus-like hole pocket gives extremal cross-section $\delta$ around $K$ and three branched, ultra small orbits $\alpha$ near $H$.
For P1, all the main extremal orbits with frequency less than 2000~T can be successfully identified by tuning $E_F$ to $+10$~meV in the calculated band structure.

\begin{figure}[htbp]
\begin{center}
\includegraphics[clip, width=0.48\textwidth]{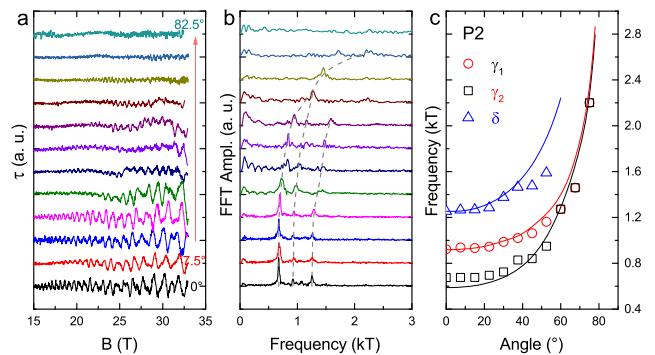}  
\caption{Color online. (a) Magnetic torque signals for P2 at different angles. $\theta$ is defined as the angle between the magnetic field direction and the crystallographic $c$ axis within the $ac$ plane. (b) Corresponding FFT spectrums. Dashed lines are a guide to the eye.  (c) FFT frequencies for the orbits $\gamma_1$, $\gamma_2$ and $\delta$ versus angle. Solid lines are calculated frequencies.}
\label{f3}
\end{center}
\end{figure}

The quasi-2D FS around $K-H$ line in the reciprocal space of PbTaSe$_2$ is verified by a systematical measurement of the torque signals at different magnetic field orientations (Fig.~\ref{f3}(a)).
The experimentally observed de Haas-van Alphen frequencies as a function of the angles are shown in Fig.~\ref{f3}(b) and (c).
The orbits $\gamma_1$,$\gamma_2$ and $\delta$ can be clearly identified at low angles and change roughly in a $1/\cos\theta$ manner.
This agrees well with the quasi-2D like feature of these bands.
It is noteworthy that the orbits $\gamma_1$ and $\gamma_2$ become merged at around $60^\circ $ and this feature matches the change of the minimal and maximal cross sections of the corrugated innermost electron tube at high angle.
The QOs become weaker above $60^\circ $ and the frequencies are hard to trace due to the appearance of other extremal cross-section areas in other pockets.

In order to extract more information of the electron and hole pockets, we now analyze the QOs observed in Fig.~\ref{f1} at different temperatures.
The temperature dependence of the QOs in torque signals at $\theta=0^\circ$ is well described by the Lifshitz-Kosevich (LK) formula\cite{shoenberg2009magnetic} as following:
$$R_T=\frac{\alpha p X}{\sinh \alpha p X}$$
where $\alpha=2\pi^2 k_B/e\hbar$, $X=m^\ast T/B$, $m^\ast$ being the effective mass. For $\gamma_1$, $\gamma_2$ and $\delta$ orbits, the fits to LK formula give $m^\ast$ as 0.42, 0.51, 0.69 $m_e$, respectively, as shown in Fig.~\ref{f4}(b).
For the oscillations in $S_{xx}$, the commonly used LK formula fails because the QOs now depend on the derivative of density of states\cite{young1973quantum,coleridge1989low}.
N. B. there is no apparent contribution to QOs from phonon-drag because the carrier density of PbTaSe$_2$ is pretty high and the overall FS is large.
Previous works pioneered by R. Fletcher \it et al.\rm  \cite{fletcher1981amplitude,coleridge1989low,fletcher1983experimental,fletcher1995oscillations,tieke1996magnetothermoelectric, morales2016thermoelectric} suggest that the thermal damping factor for diffusive part of magneto-thermopower should be
$$R_T=\frac{(\alpha pX)\coth(\alpha pX)-1}{\sinh (\alpha pX)}.$$
For the five frequencies detected in $S_{xx}$, we successfully fit their $m^\ast$ as 0.14, 0.48, 0.49, 0.67, 0.63 $m_e$ respectively, as in Fig.~\ref{f4}(a).
These values are in good agreement with those obtained in QOs in torque signals.
The $\alpha$ band possesses a light effective mass, close to those relativistic electrons observed in Weyl semimetal TaAs and TaP family\cite{PhysRevB.95.085202,zhang2016signature,zhang2017magnetic} but larger than that in Dirac semimetal Cd$_3$As$_2$\cite{he2014quantum}. For the other four orbits, including the magnetic tunneling induced orbit $\beta $, carriers are of pretty heavy effective masses around 0.5 $m_e$.

As the oscillatory frequencies for $\alpha$, $\beta$ and $\delta$ in $S_{xx}$ and $\gamma_1$ in $\tau$ are well separated from each other, the Berry phase of these orbits can be inferred by the so-called Landau fan diagram\cite{ando2013topological}.
Every peak position of $\Delta S_{xx}$ is assigned with an integral Landau index and the residual phase shifts ($\phi_s$) are shown in Fig.~\ref{f4}(e). Note for thermoelectricity, the phase shift  $\phi_s=-1/2+\phi_B+\phi_{3D}+\phi_T$, where $\phi_B$ is the Berry phase, $\phi_T$ is $1/4$ for hole carriers in thermoelectric QOs\cite{fletcher1981amplitude,coleridge1989low,fletcher1983experimental,fletcher1995oscillations,tieke1996magnetothermoelectric,matusiak2017thermoelectric,havlova1986quantum}.
The additional phase shift $\phi_{3D}$ stemming from the dispersion along $k_z$ equals $\mp 1/8$ for a maximum cross section of electron and hole pocket, respectively, but $\pm 1/8$ for a minimum cross section\cite{li2018rules}, and equals zero for a 2D sheet of Fermi surface.
For magnetic torque signals, we follow Mikitik's work on magnetization\cite{mikitik2004berry} and assign an integral value to every peak for the neck $\gamma_1$ orbit, considering the fact that $\tau=-\frac{1}{F} \frac{dF}{d\theta}M_{\parallel}HV$, where $M_\parallel$ is the parallel component of magnetization\cite{shoenberg2009magnetic}.
After all these treatments, we finally get $\phi_B$ as -0.18, 0.32, 0.01 and 0.10 for $\alpha$, $\beta$, $\gamma_1$ and $\delta$ orbits, respectively.
Detailed information about the QOs is summarized in Table~\ref{t1}.

To shed light on the experimentally observed Berry phases for these orbits, we closely investigate their origination in band structure again.
Without SOC, PbTaSe$_2$ possesses a spinless nodal ring (NL0) around $H$ point\cite{bian2016topological}.
When SOC is turned on, due to the protection of mirror symmetry, this nodal ring splits into a pair of new nodal rings (NL1 and NL2), as shown in the left panel of Fig.~\ref{f4}(c). The tiny $\alpha$ pocket arises from this gap-opening process, residing just on NL0 and interlocking with each other (Fig.~\ref{f4}(d)), which is the origin of its nontrivial topology. The $\alpha$ pocket encloses a linearly dispersive topologically nontrivial band crossing point just above $E_F$ if the small band gap ($\Delta\sim15\mathrm{meV}$) from SOC is neglected. According to Ref.~\cite{li2018rules}, a nonzero Berry phase is expected here, in the form of $\pm\frac{1}{2}(1-\frac{\Delta}{2E_F})$, where $E_F$ is the Fermi energy. As the estimated Fermi energy of $\alpha$ pocket from QOs is $10~\mathrm{meV}$, the -0.18 Berry phase corresponds to an energy gap of 13meV, in well agreement with band structure calculations.
On the other hand, the orbit $\gamma_1$ lies in the inner side of NL2 and does not interlock with it; hence this pocket is topologically trivial, as stated in Ref.~\cite{li2018rules}. The SOC also creates a third nodal ring around $K$ where $\gamma_2$ and $\delta$ orbits are located nearby. Therefore they should also display a trivial Berry phase, which is consistent with our observations.
The $\beta$ orbit stems from the magnetic breakdown of two 3D FS , which makes it hard to calculate its Berry phase directly, because an extra energy and $k_z$ dependence is introduced \cite{mikitik1999manifestation}. The 0.32 Berry phase observed deserves further calculations.

\begin{table*}[htpb]
\caption{\label{t1}
Parameters for the QOs in PbTaSe$_2$. Subscript $\tau$ denotes values obtained from torque measurements while others from thermoelectric measurements. F is the oscillatory frequency and F{\tiny cal.} is the value from DFT calculation.
}
\begin{tabular}{p{2.cm}<{\centering}|p{2.cm}<{\centering}|p{2.cm}<{\centering}|p{2.cm}<{\centering}|p{2.cm}<{\centering}|p{2.cm}<{\centering}|p{2.cm}<{\centering}|p{2.cm}<{\centering}}
\hline\hline
                Orbits      &   F (T)       & F{\tiny cal. }(T)         &  $m^\ast$ ($m_e$) &   $E_F$ ($meV$) &   Carrier type            &   $\phi_s(2\pi)$      &  $\phi_B(2\pi)$\\
\hline
                $\alpha$    &   5.8         &      6                    &   0.14(1)         &   10            &   hole                    &   -0.43(6)            &     -0.18 \\
\hline
                $\beta$     &   80.3        &      79                   &   0.48(1)         &   --            &   hole                    &   -0.80(3)            &     0.32 \\
\hline
\multirow{2}*{$\gamma_1$}   &   672         & \multirow{2}*{630}        &   0.49(6)         &   320           &   \multirow{2}*{electron} &   --                  &     --    \\
\cline{2-2}\cline{4-5}\cline{7-8}
                            &   685$_\tau$  &                           &   0.42(2)$_\tau$  &   378$_\tau$    &                           &   -0.37(3)$_\tau$     &     0.01$_\tau$\\
\hline
\multirow{2}*{$\gamma_2$}   &   910         & \multirow{2}*{925}        &   0.67(5)         &   314           &   \multirow{2}*{electron} &   --                  &      --   \\
\cline{2-2}\cline{4-5}\cline{7-8}
                            &   904$_\tau$  &                           &   0.51(5)$_\tau$  &   411$_\tau$    &                           &   --                  &      --   \\
\hline
\multirow{2}*{$\delta$}     &   1249        & \multirow{2}*{1330}       &   0.63(3)         &   460           &   \multirow{2}*{hole}     &   -0.02(2)            &     0.10  \\
\cline{2-2}\cline{4-5}\cline{7-8}
                            &   1275$_\tau$ &                           &   0.69(9)$_\tau$  &   430$_\tau$    &                           &   --                  &      --   \\

\hline\hline

\end{tabular}
\end{table*}

\section{Conclusion}

In conclusion, we observed strong QOs in magneto-thermoelectric and magnetic torque signals in PbTaSe$_2$.
We are able to trace every frequency with its complicated FS.
One frequency is related to magnetic breakdown while others are from 2D-like FS near the TNRs around the BZ corner $K-H$ line.
The angle dependence and the Berry phases extracted from QOs confirm the existence of the TNRs.
Moreover, the fine structure of the TNRs and the exact $E_F$ location is determined.
Previous studies show there exist two types of TSSs in PbTaSe$_2$. One is the drumhead surface states connecting to TNRs around $\bar{K}$, and the other is the single Dirac TSS originating from band inversion around $\bar\Gamma$ \cite{bian2016topological,chang2016topological,guan2016superconducting}.
Our observation helps better understanding these TSSs.
Our work also highlights the magneto-thermoelectric measurement for detecting the multi-frequency QOs for complicated FS.
Tracing the ultra-low frequency QOs in thermoelectric signals enables us to fathom the delicate electronic structure of multiband topological semimetals.

\section{Acknowledgement}
We would like to thank Gabriel Seyfarth and Lu Li for their constructive ideas during the process of minimizing system noises. Shuang Jia was supported by the National Natural Science Foundation of China No. U1832214, No.11774007, the National Key R\&D Program of China (2018YFA0305601) and the Strategic Priority Research Program of Chinese Academy of Sciences (Grant No. XDB28000000). Jinglei Zhang was supported by Innovative Program of Development Foundation of Hefei Center for Physical Science and Technology (2017FXCX001), Natural Science Foundation of China No. 11504378 and the Youth Innovation Promotion Association CAS (grant number 2018486). T.-R.C. was supported from Young Scholar Fellowship Program by Ministry of Science and Technology (MOST) in Taiwan, under MOST Grant for the Columbus Program MOST107-2636-M-006-004, National Cheng Kung University, Taiwan, and National Center for Theoretical Sciences (NCTS), Taiwan. This work is supported partially by the MOST, Taiwan, Grants No. MOST 107-2627-E-006-001.

%

\clearpage

\end{document}